\documentclass[preprint,showpacs,showkeys,nofootinbib]{revtex4}

\pdfoutput=1
\usepackage{color}
\newcommand{\n}{\nonumber}
\newcommand{\be}{\begin{equation}}
\newcommand{\ee}{\end{equation}}
\newcommand{\bea}{\begin{eqnarray}}
\newcommand{\eea}{\end{eqnarray}}

\begin{document}

\title{Time-dependent Darboux transformation  and \\ supersymmetric hierarchy of Fokker-Planck equations}
\author{Choon-Lin Ho}%\\
%\addresss
\affiliation{Department of Physics, Tamkang University,
Tamsui 25137, Taiwan}

%\date{2021/9/9; 12/20} 

%\maketitle  % for LaTex

\begin{abstract}

A procedure is presented for solving the Fokker-Planck equation with constant diffusion but  non-stationary drift.  It is based on the correspondence between the Fokker-Planck equation and the non-stationary Schr\"odinger equation.  The formalism of supersymmetric  quantum mechanics is extended by applying the Darboux transformation directly to the non-stationary Schr\"odinger equation. From a solution of a Fokker-Planck equation a solution of the Darboux partner is obtained. For drift coefficients satisfying the condition of shape invariance, a supersymmetric hierarchy of Fokker-Planck equations with solutions related by the Darboux transformation is obtained.

\end{abstract}

% 05.10.Gg  Stochastic analysis methods (Fokker-Planck, Langevin, etc.)
% 02.30.Ik    Integrable systems
% 05.90.+m  Other topics in statistical physics, thermodynamics, and nonlinear dynamical systems (restricted to new topics in section 05)
% 02.50.Ey   Stochastic processes


 \maketitle

%%%%%%%%%%%%%%%%%%%%%%%%%%%%%%%%%%%%

%  Section I --Introduction

%%%%%%%%%%%%%%%%%%%%%%%%%%%%%%%%%%%%

\section{Introduction}

The Fokker-Planck  equation (FPE)  is one of the basic equations used to study the effects of fluctuations in various kinds of systems
\cite{Ris,Sau}.  It has found applications in such diverse areas as physics, astrophysics, chemistry, biology, finance, etc.  Owing to
its wide applicability, various methods of finding exact and approximate solutions of the FP equations have been developed.

Generally, it is not easy to find analytic solutions of the FPE. Exact analytical solutions of the FPEs are known for only a few
cases, such as linear drift and constant diffusion coefficients \cite{Ris,Sau}.

One of the methods of solving the FPE  with time-independent drift and diffusion coefficients is to transform the FPE into a stationary Schr\"odinger equation, and then solve the eigenvalue problem of the latter.  This method is useful when the associated
Schr\"odinger equation is exactly solvable; for example, with infinite square well, harmonic oscillator potentials, etc.
Several FPEs have been exactly solved in this way \cite{Ris,Sau}. Recently, this method was employed to investigate quasi-exactly solvable  FPEs \cite{HS}, and FPEs with solutions based on the newly discovered exceptional orthogonal polynomials \cite{CH}. 

Solving the FPEs with time-dependent drift and/or diffusion coefficient is in general a difficult task. It is
therefore not surprising that the number of papers on such kind of FPE is far less than that on FPE with time-independent coefficients.  
Some recent works on the FPE with time-dependent diffusion coefficients appear in \cite{GMNT,KSF,GNT}, and works involving
time-dependent drift coefficients can be found in \cite{LM,Ho1,Ho2}. Refs. \cite{WH,OK,SS} consider the FPEs with both
time-dependent diffusion and drift coefficients. The symmetry properties of the one-dimensional FPE with arbitrary coefficients 
of drift and diffusion are investigated in \cite{SS}.  Similarity solutions of FPEs with time-dependent coefficients are considered in \cite{Ho3,Ho4,Ho5}.

For FPEs with constant diffusion coefficients and stationary drifts, the correspondence between FPE and Schr\"odinger equation allows one to exploit the formalism of supersymmetric (SUSY)  quantum mechanics \cite{SUSY}. In SUSY quantum mechanics, the  spectrum of a Schr\"odinger equation can be related to that of another through the Darboux transformation \cite{Dar,Cru}.  The SUSY method was first employed in \cite{BB} to compute  the small eigenvalue of the Fokker-Planck equation which controls the rate at which equilibrium is approached. 
A study of the solutions of FPEs by means of the strictly isospectral method of SUSY quantum mechanics was carried out in \cite{Ros}.    FPEs in periodic potentials was discussed in \cite{SMJ}.
The formalism of $n$-th order Darboux transformation given in \cite{Cru} was applied  to the FPE in \cite{SRGGR}.

More recently, FPEs with constant diffusion coefficients but non-stationary drifts was discussed in \cite{IN}. There the authors 
used the intertwining relations of SUSY Quantum Mechanics in a new  asymmetric form with a suitable ansatz to obtain a new class of analytically solvable models.

In this work we would like to consider solution of FPE with time-dependent drift and constant diffusion based on its correspondence with non-stationary Schr\"odinger equation.  The formalism of SUSY quantum mechanics is extended by applying the Darboux transformation directly to the non-stationary Schr\"odinger equation \cite{MS}. From a solution of a FPE the solution of the Darboux partner is obtained.  For drift coefficients such that the condition of shape invariance is satisfied, a SUSY hierarchy of FPEs with solutions related by the Darboux transformation is obtained.

%%%%%%%%%%%%%%%%

\section{Time-dependent Darboux transformation}

Consider the FPE with constant diffusion
\be
\frac{\partial}{\partial t} P(x,t)=\left[-\frac{\partial}{\partial x} D^{(1)}(x, t) +
\frac{\partial^2}{\partial x^2}D^{(2)}\right]P(x,t).
\label{tFPE}
\ee
The function  $P(x,t)$ describes the distribution of particles in a system.
$D^{(1)}(x,t)$ and $D^{(2)}$  are, respectively,  the drift and the diffusion
coefficient.  The drift coefficient represents the external force acting on the particle, while the diffusion coefficient accounts for the effect of
fluctuation. 
Without loss of generality, we set the diffusion constant to be unity, $D^{(2)}=1$. The drift coefficient $D^{(1)}(x,t)$ is given by a time-dependent prepotential $W(x, t)$,  $D^{(1)}(x, t)=-2 W^\prime (x, t)$, where the prime denotes the derivative with respect to $x$.  The function $W(x,t)$ is called the prepotential because, as shown below, it determines the potential of a Schr\"odinger equation related to the FPE.

Let $P(x, t)=e^{-W(x,t)}\psi(x,t)$.  From (\ref{tFPE}) we find that $\psi$ satisfies the  Schr\"odinger-like equation with 
time-dependent potential
\be
-\frac{\partial}{\partial t}\psi= 
-\frac{\partial^2}{\partial x^2}\psi 
+ \left(W^{\prime 2} - W^{\prime\prime}- {\dot W}\right)\psi, \label{S-like-1}
\ee
where ${\dot W}\equiv \partial W/\partial t$.  For time-independent drift, $ {\dot W}=0$, $\psi (x)=e^{-W(x)}$ is a solution of (\ref{S-like-1}), and hence $P(x)=e^{-2W(x)}$ is the stationary solution of the FPE.  This is not the case when ${\dot W}\neq 0$.

The Darboux transformation was discovered about 140 years ago \cite{Dar}, and later rediscovered in physics in the 80's of the last century (for a good review, see, e.g., \cite{SUSY}). 
The Darboux transformation is originally applied to time-independent Sturm-Liouville equations, of which the Schr\"odinger equation is a special case.  But it can also be applied to Schr\"odinger-like  equations with time-dependent potential such as (\ref{S-like-1}) \cite{MS}.

Consider the general equation
\be
\alpha\frac{\partial}{\partial t}\psi= 
-\frac{\partial^2}{\partial x^2}\psi 
+ V(x,t)\psi.
\label{gen}
\ee
Here $\alpha$ is a constant ($\alpha=-1$ in (\ref{S-like-1}), and $\alpha=i\hbar$ in the Schr\"odinger equation). Suppose $\psi_0(x, t)$ is a solution of this equation. 
Then it can be checked that, for any solution $\psi(x,t)$ of (\ref{gen}),  the Darboux transformed functions
 \bea
 {\tilde V}(x,t)&=&V-2\,(\ln\psi_0)^{\prime\prime},\n\\
{\tilde \psi}(x,t)&=& \left(\partial_x  -(\ln\psi_0)^\prime\right)\psi,
\label{Darboux}
 \eea
 also satisfy the same form of Schr\"odinger-like equation,
 \be
\alpha\frac{\partial}{\partial t}{\tilde \psi}= 
-\frac{\partial^2}{\partial x^2}{\tilde \psi}
+ {\tilde V}{\tilde \psi}.
\ee
The set of equations in (\ref{Darboux}) is called the (time-dependent ) Darboux transformation.

Returning to FPEs. Suppose we start with a FPE with drift coefficient defined by the prepotential $W_0 (x, t)$. Then
from (\ref{S-like-1}) the potential of the associated Schr\"odinger equation is $V=W_0^{\prime 2} - W_0^{\prime\prime}- {\dot W}_0$.  
Unlike the stationary case,  $e^{-W_0}$ is not a solution of (\ref{S-like-1}) with non-stationary $V(x,t)$.   Suppose a solution  is ${\tilde\psi}_0=e^{-{\widetilde W}_0}$. Then we have
\bea
V_0&=&W_0^{\prime 2} - W_0^{\prime\prime}- {\dot W}_0\n\\
&=&{\widetilde W}_0^{\prime 2} - {\widetilde W}_0^{\prime\prime} + {\dot {\widetilde W}}_0.
\label{R1}
\eea
Note the ``+" sign on the last term :  $+  {\dot {\widetilde W}}_0$.
This equation can be regarded as a generalized Riccati equation.  Solving (\ref{S-like-1}) with $W=W_0$ for $e^{-{\widetilde W}_0}$ is equivalent to solving for ${\widetilde W}_0$ from (\ref{R1}) given $W_0$.  This is the non-trivial part of our procedure.

By (\ref{Darboux}) the Darboux transformed potential of $V_0$ based on the state $e^{-\widetilde W_0}$ is $V_1={\widetilde W}_0^{\prime 2} + {\widetilde W}_0^{\prime\prime} + {\dot {\widetilde W}}_0$.
The presence of both positive signs in $V_1$ implies that $V_1$ does not define a Schr\"odinger equation associated with a FPE with time-dependent drift. But it may if one can find a function $W_1(x, t)$ such that
\bea
V_1(x,t)&=&{\widetilde W}_0^\prime (x, t)^2 + {\widetilde W}_0^{\prime\prime}(x, t) + {\dot {\widetilde W}}_0(x, t)\n\\
&=&W_1^\prime (x, t)^2 - W_1^{\prime\prime}(x, t ) - {\dot W}_1(x, t).
\label{R2}
\eea
This is again a generalized Riccati equation.  A simple solution of this equation is $W_1(x,t)=-{\widetilde W}_0(x,t)$.

From Eq.\,(\ref{tFPE}) and (\ref{S-like-1}),  we see that $V_1$ defines a Schr\"odinger equation that is associated with a FPE with drift $D^{(1)}=-2 W_1^\prime$.  A solution of this partner FPE can be obtained from a solution $P_0(x,t)$ of the original FPE as follows.  Let $P_0(x,t)=e^{-W_0}\psi$.  Then by  (\ref{Darboux}) the transformed function  ${\tilde \psi}=(\partial_x + {\widetilde W}_0^\prime)\psi$  satisfies the Schr\"odinger equation defined by $V_1$.  And so a solution of the partner FPE is
$P_1(x, t) =e^{-W_1} {\tilde \psi}$.  Putting all these together, we have
\bea
P_1 (x, t)&=&e^{-W_1} (\partial_x + {\widetilde W}_0^\prime)\left( e^{W_0} P_0(x, t) \right)\label{P1-1}\\
&=&e^{-W_1} (\partial_x - W_1^\prime)\left( e^{W_0} P_0(x, t) \right).\n
\eea

We note that the solution so generated may not serve as a probability density function, as it may not be normalized to be positive-definite over a suitable domain.

%---------------------------------------------------------
%%%%%%%%%%%%%%

\section{Example: Generalized Ornstein-Uhlenbeck process I}

The well-known Ornstein-Uhlenbeck process is described by a drift coefficient $D^{(1)}(x)=-\gamma x$ with a constant parameter $\gamma$.  The associated quantum system is the simple harmonic oscillator.

Let us consider the generalized Ornstein-Uhlenbeck process with $D^{(1)}(x)=-\gamma(t) x$, where $\gamma(t)$ is taken to be a function of time.  In this case the prepotential is $W_0(x,t)=\gamma(t) x^2/4$.

Eq.\,(\ref{S-like-1}) becomes
\be
-\frac{\partial}{\partial t}\psi= 
-\frac{\partial^2}{\partial x^2}\psi 
+ \left\{\frac14 \left[\gamma(t)^2-{\dot\gamma}(t)\right] x^2- \frac12\gamma(t)\right\}\psi,
\label{psi1}
\ee

Letting $\psi=e^{\int^t \gamma dt/2}\phi$, we have
\be
-\frac{\partial}{\partial t}\phi= 
-\frac{\partial^2}{\partial x^2}\phi 
+ \frac14 \left[\gamma(t)^2-{\dot\gamma}(t)\right] x^2\,\phi.\label{phi1}
\ee

As a solvable case let us choose $\gamma(t)$ such that $\gamma(t)^2-{\dot\gamma}(t)=0$, or $\gamma(t)=-1/(t+C)$ where $C$ is a real constant.  So Eq.\,(\ref{phi1}) becomes the diffusion equation
\be
\frac{\partial}{\partial t}\phi= \frac{\partial^2}{\partial x^2}\phi.
\label{phi2}
\ee
For any solution $\phi(x,t)$ of (\ref{phi2}), we have a solution
\be
P_0(x,t)=\frac{1}{\sqrt{t+C}}e^{\frac{x^2}{4(t+C)}}\phi(x,t)
\ee
for the FPE with $D^{(1)}(x)= x/(t+C)$.

Particularly, we can take $\phi(x,t)$ to be the well-known self-similar solution
\be
\phi(x,t)=\frac{1}{\sqrt{4\pi t} }e^{-\frac{~x^2}{4t}},
\ee
which gives
\be
P_0(x,t)=\frac{1}{\sqrt{4\pi t(t+C)}}e^{-\frac{C x^2}{4t(t+C)}}.
\label{uo}
\ee

Now we find a Darboux partner of this FPE by the procedure described in Sect. II.  For this we need another solution ${\tilde\psi}_0=e^{-{\widetilde W}_0}$ of (\ref{psi1}) as an auxiliary function for the Darboux transformation.    By choosing a solution of  the diffusion equation (\ref{phi2}) to be $e^{a^2 t +ax}$ ($a$ a real constant), we have
\be
{\widetilde\psi}_0=\frac{|}{\sqrt{t+C} }e^{a^2t+ax},
\ee
or,
\be
{\widetilde W}_0=\frac12 \ln (t+C) -a^2t-ax=-W_1.
\ee
So the Darboux partner FPE of the generalized Ornstein-Uhlenbeck process has a constant drift coefficient  $D^{(1)}=-2  W_1^\prime = -2a$.  

From (\ref{P1-1}), given a solution $P_0(x,t)$ in (\ref{uo}) there corresponds a solution
\be
P_1(x,t)\sim \frac{x+2at}{4 \sqrt{\pi} t^{3/2}}\,e^{-\frac{(x+2 a t )^2}{4 t}}.
\label{sol-diffuse} 
\ee
for the partner FPE.  For $a=0$, we get a solution of the diffusion equation.
 For $a>0$ ($a<0$), the drift coefficient $D^{(1)}= -2a <0$ ($>0$) represents a left (right)-pointing drift force, and (\ref{sol-diffuse}) is a solution of the corresponding FPE  that has the same shape as the solution of the $a=0$ case, but is displaced to the left (right) by a distance $2|a|t$.

%%%%%%%%%%%%

\section{Shape invariance and SUSY hierarchy of FPEs}

The prepotential $W_0(x,t)=\gamma(t) x^2/4$ in the last section satisfies the shape invariance condition
\bea
&&W_0^\prime (x, t)^2 + W_0^{\prime\prime}(x,t )\n\\
&=& W_0^\prime (x, t)^2 - W_0^{\prime\prime}(x, t) + \gamma(t).
\label{SI-osc}
\eea
This property allows us to find Darboux partners of the original FPE in an easy way.

%--------------------------

\subsection{Shape invariance}

As all the known one-dimensional exactly solvable Schr\"odinger potentials possess such property, we give here a general procedure to determine Darboux partners of FPEs associated with these potentials.  

 By shape invariance one means that the Darboux transformed potential and the original potential are similar in shape and differ only in the parameters appearing in them \cite{SUSY}. If $a_0$ is a set of parameters in $V(x; a_0)$ and ${\tilde V}(x;a_0)$,  Then mathematically shape invariance means the condition
\be
{\tilde V}(x; a_0)=V(x; a_1) + R(a_0),\n
\ee
or in terms of prepotentials,
\be
W^\prime (x; a_0)^2 + W^{\prime\prime}(x; a_0)= W^\prime (x; a_1)^2 - W^{\prime\prime}(x; a_1) + R(a_0).
\label{SI}
\ee
Here $a_1$ is a function of $a_0$, and $R(a_0)$ is an $x$-independent shift function.
  By successive application of the Darboux transformation one generates a series of potentials all similar in shape with differential parameters $a_n$'s ($n=0,1,2,\ldots$). Shape invariance turns out to be a  sufficient condition for the exact-solvability of all the well-known one-dimensional analytically solvable quantum models.  Interestingly, for the known solvable quantum systems, these parameters are related simply by a shift of constant. For example, $a_{n+1}=a_n + 1$ for the radial oscillator ($a_n$= angular momentum), and $a_{n+1}=a_n$ (unchanged) for the simple harmonic oscillator ($a_n$= angular frequency).

%------------------
\subsection{SUSY hierarchy}

Now let us  consider FPEs with drift coefficients given by the prepotentials of these solvable quantum models, but with the parameters $a_n$ changed to be  a function  of time, i.e., $a_n\to a_n(t)$.  The Ornstein-Uhlenbeck system in the previous section is just such a case. 

Suppose we start with a FPE with $D^{(1)}(x,t)=-2 W^\prime_0(x; a_n(t))$ and an initial solution $P_0(x,t)$.

The general shape invariance condition is
\bea
&& W^\prime (x,; a_n(t))^2 + W^{\prime\prime}(x; a_n(t))\label{SI-t}\\
&=& W^\prime (x; a_{n+1}(t))^2 - W^{\prime\prime}(x; a_{n+1}(t)) + R(a_n(t)).
\n
\eea
The generalized Riccati equation (\ref{R1}) can be recast into
\bea
V&=&W^\prime (x,; a_n(t))^2 - W^{\prime\prime}(x; a_n(t)) - {\dot W}_0(x; a_n(t))\n\\
&=& W^\prime (x; a_{n-1}(t))^2 + W^{\prime\prime}(x; a_{n-1}(t)) \n\\
&& -  R(a_{n-1}(t)) - {\dot W}_0(x; a_n(t))\\
&=& {\widetilde W}_0^{\prime 2}(x,t) - {\widetilde W}_0^{\prime\prime} (x,t)+ {\dot {\widetilde W}}_0(x,t).\n
\eea
A solution ${\widetilde W}_0$ can be taken as
\be
{\widetilde W}_0 (x,t)= - W_0(x; a_{n-1}(t)) -  \int^t R(a_{n-1}(t))\,dt,
\ee
and a solution of (\ref{R2}) can be chosen to be 
\bea
&&W_1 (x,t)=-{\widetilde W}_0 (x,t)\n\\
=&&  W_0(x; a_{n-1}(t)) + \int^t R(a_{n-1}(t))\,dt,
\eea
as long as we have
\be
{\dot W}_0(x; a_{n-1}(t))={\dot W}_0(x; a_n(t)).
\label{W0}
\ee
In view of the fact that $a_n(t)$ and $a_{n-1}(t)$ differ only by a constant, the condition (\ref{W0}) is true if $a_n(t)$ appears in $W_0(x,t)$ only as a multiplicative factor of a function of $x$.  Of the ten solvable one-dimensional quantum models, such condition is true for: the 1d oscillator, the 3d oscillator, the Morse, the Scarf I and II, and the P\"oschl--Teller potential.

The Darboux partner is then the FPE with $D^{(1)}=-2W_0^\prime (x; a_{n-1}(t))$,  and a solution $P_1(x,t)$ is given by (\ref{P1-1}).

This process can be iterated, and generates from the solution $P_0(x,t)$ of the FPE with $D_0^{(1)}=-2 W_0^\prime (x,a_n(t))$ a solution $P_k(x,t)$ of the FPE with drift $D_k^{(1)}=-2 W_0^\prime (x; a_{n-k}(t))$ ($k=1, 2, \dots$), up to the lowest $a_n$ allowed by the model,
\bea
P_k(x, t)  &=&e^{-W_k }\left(\partial_x - W_k^\prime\right) \label{P1-2} \\
&& \times\left( e^{W_{k-1}} P_{k-1}(x, t) \right),~~k=1,2,\ldots\n
\eea
where we have employed the abbreviation
\bea
&&W_k \equiv W_k(x,t)\n\\
&=& W_0(x, a_{n-k}(t))+ \int^t \sum_{s=n-k}^{n-1}\,R(a_s(t))\,dt.
\eea

We shall call these FPEs, related by the Darboux transformation,  the SUSY hierarchy of FPEs.
%--------------------------------------------

\subsection{Example: Generalized Ornstein-Uhlenbeck II}

We take the generalized Ornstein-Uhlenbeck process in Sect.\,III as an example to illustrate the procedure. Here $W_0(x,t)=\gamma(t)x^2/4$. This prepotential has the distinctive characteristics, as is evident from (\ref{SI-osc}),  that the potential defined by it is exactly the same under the Darboux transformation, except by an $x$-independent shift $R(a_n(t))$, i.e., the parameters $a_n=\gamma(t)$ are all the same, and $R(a_n(t))=\gamma(t)$.  This means all the solutions $P_k(x,t)$ in the SUSY hierarchy are the solutions of the same FPE with $D^{(1)}=- \gamma(t) x$.

For $\gamma(t)=-1/(t+C)$, we get from $P_0(x,t)$ in (\ref{uo}) the next two solutions in the hierarchy as

\be 
P_1(x,t)\propto \frac{Cx}{4 t \sqrt{\pi t(t+C)}}\,e^{-\frac{C x^2}{4 t (t+C)}},
\ee
 and
\be 
P_2(x,t)\propto \frac{C (C x^2-2 C t -2 t^2 )}{8 t^2 \sqrt{\pi t(t+C)}}\,e^{-\frac{C x^2}{4 t (t+C)}}.
 \ee
 
 %%%%%%%%%%%%%%%%%%%%%%%%%%%%%%%%%%%%%%%%%%%%

\section{FPEs with time-independent drifts}

\subsection{Backward hierarchy}

If the parameters $a_n$'s in Sect. IV are independent of time, we obtain solutions of FPEs with time-independent drifts which involved the  parameters $a_n$ related by shape invariance.

\be
W_k \equiv W_k(x)= W_0(x; a_{n-k})+ t \sum_{s=n-k}^{n-1}\,R(a_s),~~k=1,2,\ldots
\ee
Eq.\,(\ref{P1-2}) becomes
\bea
&&P_k (x, t) =e^{-W_0(x; a_{n-k}) - R(a_{n-k}) t }\label{B}\\
&&\times (\partial_x - W_0^\prime(x; a_{n-k}))\left( e^{W_0(x; a_{n-k+1})} P_{k-1}(x, t) \right).\n
\eea
We have ignored the integration constant  in ${\widetilde W}_1 (t)$, as it contributes only as a multiplicative constant in the expression of $P_k (x,t)$.

These FPEs are generated successively in the direction of decreasing  the index of $a_n$, so we call them the backward SUSY hierarchy of FPEs.  Below we show that there is a forward hierarchy in that the FPEs are generated by increasing the index of $a_n$.

%--------------------------------------------
\subsection{Forward hierarchy}

For FPE with time-independent drift, one can treat it as a special case of the FPE with time-dependent drift discussed previously.

Let us consider the case, namely, the prepotential $W_0(x,t)$ is separable into a sum of a space-dependent and a time-dependent part, i.e., $W_0(x, t)=W_0(x) + W_0(t)$.  Such prepotential determines a FPE  with stationary drift, as the drift coefficient $D^{(1)}(x,t)=- 2W_0^\prime (x, t)=-2 W_0^\prime (x)$ is independent of time.  The time-dependent part $W_0(t)$ indicates the arbitrariness in defining a time-independent drift using the prepotential, much like the arbitrariness in defining a mechanical force by the potential. We shall fix $W_0(t)$ by hand for the initial FPE, e.g., $W_0(t)=0$, and determine that of the partner system by solving (\ref{R2}).

In this case, the prepotential ${\widetilde W}_0 (x, t)=W_0(x) - W_0(t)$ solves Eq.\,(\ref{R1}), i.e., $e^{-{\widetilde W}_0 (x, t)}$ is a solution of (\ref{S-like-1}) with $W=W_0 (x,t)$.  To solve (\ref{R2}), we assume also that $W_1 (x,t)$ is separable, $W_1 (x,t)=  W_1 (x) + W_1 (t)$. Then (\ref{R2}) gives
\bea
&&W_0^\prime (x)^2 + W_0^{\prime\prime}(x) -  {\dot W}_0(t)\n\\
&=&W_1^\prime (x)^2 - W_1^{\prime\prime}(x) - {\dot W}_1(t).
\label{R3}
\eea

Now let us  choose $W_0(x)=W_0(x; a_n)$ to be one of those prepotentials that satisfy the shape-invariant condition.  Eq.\,(\ref{R3}) becomes
\bea
&&W_0^\prime (x; a_n)^2 + W_0^{\prime\prime}(x; a_n) -  {\dot W}_0(t)\label{R3-1}\\
&=&W_0^\prime (x; a_{n+1})^2 - W_0^{\prime\prime}(x; a_{n+1}) + R(a_n) -  {\dot W}_0(t),\n
\eea
where we have made explicit the dependence of parameters $a_n$ and $a_{n+1}$ in the prepotentials.
Comparing (\ref{R3}) and (\ref{R3-1}), we may take as solutions
$W_1 (x)= W_0 (x; a_{n+1}), ~~ W_1 (t)=W_0(t) - R(a_n) t$.
Again we have ignored the integration constant  in $W_1 (t)$, as it contributes only as a multiplicative constant in the expression of $P_1 (x,t)$.

Finally, putting  $W_0(x,t)$ (with the initial choice  $W_0(t)=0$) and $W_1 (x,t)$ in (\ref{P1-1}), we get
\bea
&&P_1 (x, t)=e^{-(W_0(x; a_{n+1})- R(a_n) t)} \\
&&\times \left(\partial_x + W_0^\prime (x;  a_n)\right)\left( e^{W_0 (x; a_n)} P_0(x, t) \right)\n
\eea
as a solution of the FPE with $D^{(1)}(x)=-2 W_0^\prime (x; a_{n+1})$.
Iterating the process, we obtain a  solution
\bea
&&P_k (x, t)=e^{-(W_0(x; a_{n+k})- R(a_{n+k-1}) t)}\label{F}\\
&& \times \left(\partial_x + W_0^\prime (x;  a_{n+k-1})\right)\left( e^{W_0 (x; a_{n+k-1})} P_{k-1}(x, t) \right)\n
\eea
of the FPE with $D^{(1)}(x)=-2 W_0^\prime (x; a_{n+k})$.

In these two subsections we have obtained the forward and  backward hierarchies of FPEs with stationary drifts by treating these FPEs as special cases of the FPEs considered in Sect.\,IV.  They can also be directly derived from the usual SUSY quantum mechanics.  This we outline in the Appendix.

%-----------------------------------------------------------------

\subsection{Ornstein-Uhlenbeck Process}

We  illustrate the results here using the original Ornstein-Uhlenbeck process with a constant drift coefficient $D^{(1)}(x)=-\gamma x$.  In this case  $a_n=R(a_n)=\gamma$ for all $n$'s.  As in Sect.\,IV.C, the drift coefficient remain the same under the Darboux transformation, and so the solutions in the SUSY hierarchy are the solutions of the same FPE. 

Let us denote by $P_{\pm k}(x,t)$ ($k=0,1,2,\ldots$) the solutions of FPEs in the forward ($+k)$ and the backward ($-k)$ SUSY hierarchy, respectively.   Then the expressions (\ref{B}) and (\ref{F}) can be unified as
\bea
&&P_{\pm (k+1)} (x, t)\n\\=
&&e^{-\frac14 \gamma x^2 \pm \gamma t} \left(\partial_x \pm  \frac12 \gamma x\right)\left( e^{\frac14 \gamma x^2} P_{\pm k}(x, t) \right),\n\\
&& ~~~~~~~  k=0,1,2,\ldots
\eea

As the initial member  $P_0(x,t)$ we take the well-known Ornstein-Uhlenbeck solution 
\be
P_0(x,t)=\sqrt{\frac{\gamma}{2\pi (1-e^{-2\gamma t})}}\,e^{-\frac{\gamma x^2}{2 (1-e^{-2\gamma t})}},
\label{ou1}
\ee
with the  initial profile $P_0(x,0)=\delta(x)$.

The first two members in the forward and backward hierarchies are:
\bea
&&P_1(x,t)=P_{-1}(x,t)\n\\
\propto &&
\frac{\gamma x}{e^{2\gamma t}-1}\,\sqrt{\frac{\gamma}{2\pi (1-e^{-2\gamma t})}}\,e^{-\frac{\gamma x^2}{2 (1-e^{-2\gamma t})}+\gamma t},\n\\
P_2(x,t)&\propto& 
\frac{\gamma (1+\gamma x^2- e^{2\gamma t})}{(e^{2\gamma t}-1)^2}\\
&&\times\sqrt{\frac{\gamma}{2\pi (1-e^{-2\gamma t})}}\,e^{-\frac{\gamma x^2}{2 (1-e^{-2\gamma t})}+2 \gamma t},\n\\
P_{-2}(x,t)&\propto& 
\gamma\frac{1+(\gamma x^2-1) e^{2\gamma t}}{(e^{2\gamma t}-1)^2}\n\\
&&\times\sqrt{\frac{\gamma}{2\pi (1-e^{-2\gamma t})}}\,e^{-\frac{\gamma x^2}{2 (1-e^{-2\gamma t})}},\n
\eea

We note that the first member of the forward and the backward hierarchy, $P_{\pm 1}(x,t)$, are the same.   This can be easily verified from the eigen-expansion of (\ref{ou1}) (which is in terms of the Hermite polynomials), and the actions of the operators $(\partial_x \pm  \gamma x/2)$ on $P_0(x,t)$.

%%%%%%%%%%%%%%%%%

\section{Summary}

In this work have considered solution of FPE with time-dependent drift and constant diffusion based on its connection with the non-stationary Schr\"odinger equation.  The Darboux transformation is directly applied to the non-stationary Schr\"odinger equation.  From a solution of a FPE a solution of the Darboux partner is derived. For drift coefficients such that the condition of shape invariance is satisfied, a SUSY hierarchy of FPEs with solutions related by the Darboux transformation is obtained.  The solution so obtained solves the FPE, but may not represent probability density as it may not be positive-definite over a certain domain.

The main part of the procedure presented here is to solve the two generalized Riccati equations, (\ref{R1}) and (\ref{R2}).  Unlike the original  Riccati equation, these two generalized equations are partial differential equations involving both the time and the space derivatives. Solving (\ref{R1}) is equivalent to solving the FPE (\ref{tFPE}) for a solution, which is in general non-trivial.  For Eq.\,(\ref{R2}), we have taken the trivial solution 
$W_1(x,t)=-{\widetilde W}_0(x,t)$ in this work.   It would be interesting to investigate the general solutions of these generalized Riccati equations.

% -----------------------------------------------------------------------------------------------------------------------------------------------------------------------------------------------

\acknowledgments

The work is supported in part by the Ministry of Science and Technology (MoST)
of the Republic of China under Grant MOST 109-2112-M-032-008 and MOST 110-2112-M-032-011.

%---------------------------------

%%%%%%%%%%%%%%%%%%%%%%%

\appendix*

\section{SUSY hierarchy of FPEs with stationary drifts}

We outline the derivation of forward SUSY hierarchy of FPEs with stationary drifts based on SUSY quantum mechanics.
The backward hierarchy can be obtained in a similar way.

%______________________________

The FPE is
\be
\frac{\partial}{\partial t} P(x,t)=\left[-\frac{\partial}{\partial x} D^{(1)}(x) +
\frac{\partial^2}{\partial x^2}\right]P(x,t).
\label{FPE}
\ee
Again we let $P(x, t)=e^{-W(x)}\psi(x,t)$
for time-independent drift $D^{(1)}(x)=-2W^\prime(x)$.
Eq.\,(\ref{S-like-1}) becomes
\begin{equation}
-\frac{\partial}{\partial t}\psi= 
-\frac{\partial^2}{\partial x^2}\psi 
+ \left(W^{\prime 2} - W^{\prime\prime}\right)\psi. \label{S-like-2}
\end{equation}
Letting $\psi(x,t)=\exp(-\lambda t)\phi(x)$, we obtain
\begin{equation}
H\phi=\lambda\phi,
\label{H}
\end{equation}
where
\begin{eqnarray}
H&\equiv&-\frac{\partial^2}{\partial x^2}+V(x),\label{V}\\
V(x)&=&W^\prime (x)^2 - W^{\prime\prime}(x).\n
\end{eqnarray}

Thus $\phi$ satisfies the stationary Schr\"odinger equation with Hamiltonian $H$ and eigenvalue $\lambda$, and $\phi_0\equiv \exp(-W)$ is the zero
mode of $H$: $H\phi_0=0$. Since only derivatives of $W(x)$ appear in $V(x)$, $W(x)$ is defined only up to an additive constant.  We
choose the constant in such a way as to normalize $\phi_0(x)$ properly, $\int\phi_0(x)^2\,dx=1$. For simplicity of presentation, we consider the cases in which the ground state wave functions are square integrable, that is the corresponding FP operators have the normalizable stationary distribution.

Now comes the essence of the method of eigenfunction expansion for solving FPEs. If all the eigenfunctions $\phi_n$
($n=0,1,2,\ldots$) of $H$ with eigenvalues $\lambda_n$ are solved, then the solution of  $P_n(x,t)$  of (\ref{FPE}) corresponding to
the eigenvalue $\lambda_n$ is $P_n(x,t)=\phi_0(x)\phi_n(x)\exp(-\lambda_n t)$.  The stationary distribution is $P_0(x)=\phi_0^2=\exp(-2W)$, which is
obviously non-negative. Any positive definite initial probability density $P(x,0)$ can be expanded as $P(x,0)=\phi_0(x)\sum_n c_n\phi_n(x)$, with constant coefficients $c_n$ ($n=0,1,\ldots$)
\begin{eqnarray}
c_n=\int_{-\infty}^\infty \phi_n(x)\left(\phi_0^{-1}(x)
P(x,0)\right)dx.
\end{eqnarray}
Then at any later time $t$, the solution of the FP equation is
\be
P(x,t)=\phi_0(x)\sum_n c_n \phi_n(x)\exp(-\lambda_n t).
\label{sol}
\ee

 Suppose $\phi(x)$ and $\phi_k(x)$, corresponding to eigenvalues $\lambda$ and $\lambda_k$, respectively,  satisfy the Schr\"odinger equation Eq.\,(\ref{H}).  Darboux theorem states that 
 the set of functions defined by the following transformations,
 \bea
 {\widetilde V}&=&V(x)-2(\ln\phi_k)^{\prime\prime},\n\\
{\tilde \phi}&=& A_0\phi\equiv \left(\partial_x  -(\ln\phi_k)^\prime \right)\phi,
 \eea
 also satisfy the same form of Schr\"odinger equation with the same eigenvalue $\lambda$,
 \be
\left( -\frac{\partial^2}{\partial x^2}+{\tilde V}(x)\right) {\tilde\phi}(x)=\lambda {\tilde\phi}(x).
\ee
 
Let us take $k=0$, and so $\lambda_0=0$ and $\phi_0(x)=e^{-W(x)}$ is the ground state of Eq.\,(\ref{H}). Then the set of functions 
$\{A_0\phi_1, A_0\phi_2, A_0\phi_3,\ldots\}$  ($A_0=\partial_x +W^\prime$) is the set of eigenfunctions of the Schr\"odinger equation with potential 
\be
{\widetilde V}=W^\prime (x)^2 + W^{\prime\prime}(x),
\label{tV}
\ee
corresponding to the eigenvalues $\{\lambda_1, \lambda_2, \lambda_3,\ldots\}$.
Thus the two Schr\"odinger equations have the same spectrum, except the ground $\lambda_0$ since $A_0\phi_0=0$. The ground state of the Darboux transformed system is at the same energy $\lambda_1$ as the first excited state of the original system.

It is customary to label the eigen-energies and eigenstates by indices $n=0,1,2,\ldots$. So we denote the energies and states of the transformed system by ${\tilde\lambda}_n$ and ${\tilde\phi}_n$, $n=0, 1,2,\lambda$. Hence we have
\bea
{\tilde\lambda}_n&=&\lambda_{n+1},\ \ \ \ \ {\tilde\phi}_n=A_0\phi_{n+1} ~~  ({\rm unnormalized}),\n\\
&&~~~~~ n=0,1,2,\ldots.
\label{link}
\eea

Up to now we only discuss the relation between two partner Schr\"odinger equations through Darboux transformation.  Now we turn to FPE. 
We know the Schr\"odinger equation (\ref{H}) is related to the FPE (\ref{FPE}).  Could the transformed Sch\"odinger equation be related to a FPE? The positive sign on the r.h.s. of (\ref{tV}) indicates that it is not --- as seen from Eq.(\ref{V}) a negative sign is needed in the potential.

However, if $V$ and $\tilde V$ are shape-invariant, then the partner Schr\"odinger equation can  be related to a FPE.  By shape invariance one means that the two potentials are similar in shape and differ only in the parameters appearing in them \cite{SUSY}. If $a_0$ is a set of parameters in $V(x; a_0)$ and ${\tilde V}(x;a_0)$,  then mathematically shape invariance means the condition
\be
{\widetilde V}(x; a_0)=V(x; a_1) + R(a_0),\n
\ee
or in terms of prepotentials,
\be
W^\prime (x; a_0)^2 + W^{\prime\prime}(x; a_0)= W^\prime (x; a_1)^2 - W^{\prime\prime}(x; a_1) + R(a_0).
\label{SI}
\ee
Here $a_1$ is a function of $a_0$, and $R(a_0)$ is $x$-independent.
Shape invariance turns out to be a  sufficient condition for the exact-solvability of all the well-known one-dimensional analytically solvable quantum models.

From (\ref{SI}) it is seen that the potential  ${\bar V}(x; a_1)\equiv  W^\prime (x; a_1)^2 - W^{\prime\prime}(x; a_1)$ defines a Schr\"odinger equation that is associated with a FPE (\ref{FPE}) with a drift  coefficient $D^{(1)}(x)=-2 W^\prime (x ;a_1)$.  We shall call these two FPEs Darboux partners.

From now on, let us use the notations $W_0\equiv W(x; a_0)$, $W_1\equiv (x; a_1)$, etc., for the sake of simplicity of mathematical expressions.  So $V=W_0^{\prime 2}- W_0^{\prime\prime}$ and ${\bar V}=W_1^{\prime 2}- W_1^{\prime\prime}$.

The ground state of potential $\bar V$ is $\bar \phi=e^{-W_1}$ with energy ${\bar \lambda}=0$. But from the point of view of the potential $V$, it has energy of the first excited state of $V$, namely, $\lambda_1=R(a_0)$.  The relations between the energies $\{\lambda_n, {\bar\lambda}_n\}$ and eigenstates $\{\phi_n, {\bar\phi}_n\}$ of the partner potentials $\{V, {\bar V}\}$, Eq.\,(\ref{link}), is rewritten as
\bea
{\bar\lambda}_n&=&\lambda_{n+1}- R(a_0),\ \ \ \ \ {\bar\phi}_n=A_0\,\phi_{n+1}
 ~~({\rm unnormalized}),\n\\
~~ &&A_0=\partial_x +W_0^\prime,~~~~~ n=0,1,2,\ldots.
\label{link1}
\eea

Obviously, for any two arbitrary constants $a_0$ and $a_1$, we can always obtain from a solution $P_0(x, t;  a_0)$ of the FPE with $D^{(1)}(x)=-2W_0^\prime$ a solution $P_1(x, t; a_1)=P_0(x,t; a_1)$ of the FPE with $D^{(1)}(x)=-2W_1^\prime$ by simply replacing $a_0$ with $a_1$. Particularly, the stationary solution $P_0(x; a_0)=\phi_0(x)^2=e^{-2 W_0}$ is mapped into $P_1(x; a_1)={\bar\phi}(x)^2=e^{-2 W_1}$. 

What we would like to point out is that, if $a_0$ and $a_1$ are not arbitrary but are related by the shape invariance condition, then
we can get yet another solution of the second FPE from that of the first as follows.

Let us relabel $P(x,t)$ in (\ref{sol}) by $P_0(x, t)$. We indicate dependence on $a_0$ by the subscript.  Consider 
$\psi(x,t)=\phi_0^{-1} P_0(x,t)=\sum_ {n=0} ^\infty\,c_n\phi_n(x)\exp(-\lambda_n t)$.  Applying $A_0$ on $\psi$ gives,
on using (\ref{link1}),
\bea
&&A_0\left(\phi_0^{-1} P_0(x,t)\right) =\sum_ {n=0} ^\infty\,c_n\left(A_0\phi_n\right)\exp(-\lambda_n t)\n\\
&=& e^{-R_0 t}  \sum_ {n=0} ^\infty\,c_{n+1} {\bar\phi}_n \exp(-{\bar\lambda}_n t),
\label{AP}
\eea
where $R_0\equiv R(a_0)$ using our convention.  Eq.\,(\ref{AP}) is a linear superposition of ${\bar\phi}_n$'s with the coefficients $\{c_1, c_2, \ldots\}$ from $P_0(x,t)$.

From (\ref{sol}) and (\ref{AP}), we see that a solution of the partner FPE is
\be
P_1(x,t)={\bar\phi}_0 \sum_ {n=0} ^\infty\,c_{n+1} {\bar\phi}_n \exp(-{\bar\lambda}_n t),
\ee
or
\bea
P_1(x,t) &=& {\bar\phi}_0 e^{R_0 t} A_0\left(\phi_0^{-1} P_0(x,t)\right)\\
&=& e^{-(W_1 - R_0 t)} \left(\partial_x +W_0^\prime\right) \left(e^{W_0}\, P_0(x,t)\right).\n
\eea

This procedure can be continued.   The shape invariance condition gives  rise to a sequence of parameters $\{a_0, a_1, a_2, \dots\}$. 
Each set of parameters $a_n$ defines a FPE in the SUSY hierarchy with $D^{(1)}=-2 W_n^\prime$ ($W_n\equiv W(a_n)$).  And from a solution  $P_n(x,t)$ of one member with $a_n$, we get a Darboux-related solution $P_{n+1}(x, t)$ of the next member  by
\be
P_{n+1}(x,t) 
= e^{-(W_{n+1} - R_n t)} \left(\partial_x +W_n^\prime\right) \left(e^{W_n}\, P_n(x,t)\right),
\label{P1-gen}
\ee
which is (\ref{F}) with relabeling of indices.

%%%%%%%%%%%%%%%%%%

%%%%%%%%%%%%%

%-----------------------------
\end{document}